\documentclass[reprint,amsmath,amssymb,aps,pra,superscriptaddress, showpacs, eprint]{revtex4-1}
\usepackage[utf8]{inputenc}

\usepackage{amsmath,amssymb,ascmac,fancybox}
\usepackage{color}
\usepackage{graphicx}
\usepackage{url}
\usepackage{siunitx}
\usepackage{bm}
\usepackage{physics}
\usepackage{mathtools}
\usepackage{hyperref}
\usepackage[T1]{fontenc}
\usepackage{ulem}
\graphicspath{figures/}

\newcommand{\ene}{\varepsilon}
\renewcommand{\vec}{\vb*}

\newcommand{\re}{\mathrm{rea}}
\newcommand{\ab}{\mathrm{abs}}

\newcommand{\manifsm}{\underbar{$G$}}

\newcommand{\manicurv}{\underbar{$F$}}

\newcommand{\sigmaS}{\sigma^{L}}
\newcommand{\sigmaH}{\sigma^{H}}

\newcommand{\proj}{{P}}

\newcommand{\gqmet}{\underbar{$\mathcal{Q}$}}

\newcommand{\MoTe}{MoTe$_2$}

\begin{document}

\date{\today}
\title{Fundamental bound on topological gap} 
\author{Yugo Onishi}
\affiliation{Department of Physics, Massachusetts Institute of Technology, Cambridge, MA 02139, USA}
\author{Liang Fu}
\affiliation{Department of Physics, Massachusetts Institute of Technology, Cambridge, MA 02139, USA}

\begin{abstract}
We provide a universal tight bound on the energy gap of topological insulators by exploring relationships between topology, quantum geometry, and optical absorption. Applications of our theory to infrared absorption near topological band inversion, magnetic circular dichorism in Chern insulators, and topological gap in moir\'e materials are demonstrated.

\end{abstract}

\maketitle

A fundamental property of all insulating states of matter is the presence of an energy gap, the minimum amount of energy that can be absorbed by the system. Insulating states are not all equivalent, and can be distinguished by the topological property of the ground state wavefunction. In particular, topologically nontrivial insulators, such as Chern insulators \cite{Thouless1982, haldane_model_1988}, cannot be smoothly connected to trivial atomic insulators without closing the energy gap. 

In this work, we ask the question: is there a fundamental bound on the energy gap of topological insulators? And we provide an affirmative answer. A general tight bound on the energy gap of topological insulators having finite  Chern numbers is derived by relating topological invariant to optical absorption. Our work also reveals a direct connection between quantum geometry and optical absorption.

Our bound on topological gap can be derived in a surprisingly simple way as follows. Consider the optical conductivity $\sigma(\omega)$ of an insulator with energy gap $\Delta$. The optical conductivity consists of longitudinal and Hall parts $\sigma^L$ and $\sigma^H$, $\sigma(\omega) = \sigma^L(\omega) + \sigma^H(\omega)$, which are symmetric and antisymmetric with respect to spatial indices respectively. In particular, $\Re\sigma^L$ represents the absorption of linearly polarized light, while the imaginary part $\Im\sigma^H$ represents magnetic circular dichroism, namely, the difference in the absorption of left-handed and right-handed circularly polarized light. 

Let us now consider the absorbed power $P_{\pm}$ under circularly polarized light of left and right handedness 
${\bm E}(t) = E \left( \cos (\omega t),   \pm \sin (\omega t)  \right)$, which is given by
\begin{eqnarray}
P_{\pm} =  \left(\Re \sigma_{xx}  \pm   \Im \sigma_{xy}  \right) E^2.  \label{eq:absorbed_power}
\end{eqnarray} 
(For now we assume isotropic optical conductivity to simplify the notation; the general proof will be presented later.)   
Since $P_\pm$ must be non-negative for any frequency, we must have 
\begin{align}
    \Re\sigma_{xx} \ge \abs{\Im\sigma_{xy}}. \label{eq:abs_inequality}
\end{align}
It then follows that 
\begin{eqnarray}
 \int^{\infty}_{0} d\omega  \frac{\Re \sigma_{xx}(\omega)}{\omega} &\ge & 
\int^{\infty}_{0} d\omega  \frac{ \abs{ \Im \sigma_{xy}(\omega)}}{\omega} \nonumber \\
&\ge &  \abs{ \int^{\infty}_{0} d\omega  \frac{ \Im \sigma_{xy}(\omega)}{\omega} }. \label{eq:gos_inequality}
\end{eqnarray} 

To proceed,  we apply the Kramers–Kronig relation to relate the real and imaginary parts of the optical Hall conductivity: 
\begin{align}
    \int_0^{\infty} \dd{\omega} \frac{\Im\sigma_{xy}(\omega)}{\omega} &= -\frac{\pi}{2}\sigma_{xy}(0). \label{eq:KK}
\end{align}
The dc Hall conductivity $\sigma_{xy}(0)$ is purely real. For insulators, $\sigma_{xy}(0)$ is equal to the quantized many-body Chern number of the ground state~\cite{Thouless1982}:  
\begin{align}
    \sigma_{xy}(0) &= \frac{e^2}{h} C, \label{eq:TKNN}\end{align}
where $C$ takes integer or fractional values for integer or fractional Chern insulators respectively~\cite{niu_quantized_1985}.  

Combining the inequality (\ref{eq:gos_inequality}) with the Kramers-Kronig relation Eq. (\ref{eq:KK}) and the TKNN formula (\ref{eq:TKNN}), we find
\begin{eqnarray}
 \int^{\infty}_{0} d\omega  \frac{\Re \sigma_{xx}(\omega)}{\omega} \ge 
\frac{e^2}{4\hbar } \abs{C}. 
\end{eqnarray} 
Since optical absorption can only occur at frequencies above the energy gap $\Delta$, the left hand side is always smaller than or equal to $\hbar \int^{\infty}_{0} d\omega  \Re \sigma_{xx}(\omega) /\Delta \equiv \hbar W^0/\Delta$, where $W^0$ is the optical spectral weight. Hence we find the following upper bound on the energy gap:
\begin{align}
    \Delta \le \frac{4 \hbar W^0}{e^2  \abs{C} }. \label{eq:gapbound_intro}
\end{align}

Finally, using the standard optical sum rule that relates the spectral weight to charge density $n$ and mass $m$ (more discussions later)  
\begin{eqnarray}
   W^0 =  \frac{\pi}{2} \frac{n e^2}{m}, 
\end{eqnarray}
we arrive at a simple and elegant result:  
\begin{align}
    \Delta \le \frac{2\pi \hbar^2 n}{m\abs{C}}. \label{eq:gapbound_intro}
\end{align}
This relation sets the fundamental bound on the energy gap on Chern insulators. As seen from the above derivation, this result holds in complete generality and is even applicable to strongly interacting systems such as fractional Chern insulators.

In the following sections, we elaborate on how this gap bound can be understood in terms of the topology, quantum geometry, and energy absorption.
We begin by presenting a new sum rule for optical absorption and magnetic circular dichroism in insulators. Unlike the standard $f$-sum rule that relates the optical spectral weight to the charge stiffness~\cite{Kubo1957a}, our sum rule relates a generalized optical weight---defined as the negative-first moment of the absorptive part of optical conductivity--- %
to the quantum geometry of occupied and excited bands, see Sec.~\ref{sec:optical_weight} and ~\ref{sec:geometry_fsum_rule}. When the full frequency range $0<\omega<\infty$ is considered, the generalized optical weight is shown to be a ground state property, and the real part of our sum rule recovers a relation previously derived in the study of electronic polarization in insulators \cite{souza_polarization_2000}. 

The quantum geometric description of optical conductivity enables us to connect topological invariant and the energy gap.   
Using the standard and generalized optical sum rules, we discover a general tight bound on the energy gap of topological insulators  (Sec.~\ref{sec:weight_invariant_dichroism} and ~\ref{sec:weight_gap}).  Applications of our theory to infrared absorption near topological band inversion (Sec.~\ref{sec:infrared_absorption}), magnetic circular dichroism in Chern insulators (Sec.~\ref{sec:weight_invariant_dichroism}), and topological gap in moir\'e bands in twisted semiconductor bilayers (Sec.~\ref{sec:topological_gap_bound}) are demonstrated. Although the most of the discussion is done for noninteracting systems for simplicity, the topological gap bound is carried over to interacting systems, as we can see from the above derivation. We will discuss the generalization of our results to interacting systems in Sec.~\ref{sec:generalization_interacting}.

\section{Generalized optical weight} \label{sec:optical_weight}
With the Kubo formula, we can calculate $\sigma_{\mu\nu}(\omega)$ for general noninteracting electronic systems. 
In the present work, we shall mostly consider the optical conductivity of insulators.
Then, the optical conductivity $\sigma_{\mu\nu}(\omega)$ at zero temperature is determined by interband transitions as given by: 
\begin{align}
    \sigma_{\mu\nu}(\omega)  
    &= \frac{e^2}{\hbar} \int[\dd{\vec{k}}] \sum_{a,b} \frac{-i\ene_{ab}A^{\mu}_{ab}A^{\nu}_{ba}}{\hbar\omega+\ene_{ab}+i\delta} f_{ab}, \label{eq:optical_conductivity}
\end{align}
where $a, b$ are indices for the bands, $\ene_{ab}=\ene_{a}(\vec{k})-\ene_b(\vec{k})$, $\ene_a(\vec{k})$ is the band dispersion. $f_{ab}=f_a-f_b$ with $f_a = \Theta(\mu-\ene_a(\vec{k}))$ the Fermi distribution function at zero temperature with the chemical potential $\mu$. $e(<0)$ and $\hbar$ are the charge of electrons and the Planck constant. The integral is over the Brillouin zone, and $[\dd{\vec{k}}]$ is shorthand for $\dd^d{\vec{k}}/(2\pi)^d$ with the spatial dimension $d$. $\delta$ is an infinitesimal positive quantity appearing in the Kubo formula. 
$\vec{A}_{ab}$ is the interband Berry connection defined as 
\begin{align}
    \vec{A}_{ab} &= \mel{u_{\vec{k}a}}{i\nabla_{\vec{k}}}{u_{\vec{k}b}},
\end{align}
where $\ket{u_{\vec{k}a}}$ is the cell-periodic part of Bloch wavefunction of the band $a$ at wave vector $\vec{k}$. In deriving Eq.~\eqref{eq:optical_conductivity}, we neglect the wavevector of light and coupling to the magnetic field. 

The optical conductivity \eqref{eq:optical_conductivity} can be separated into the symmetric part ($\sigmaS$) and the antisymmetric part ($\sigmaH$) with respect to spatial indices:
\begin{align}
    \sigma^{L,H}_{\mu\nu}(\omega) = \frac{\sigma_{\mu\nu}(\omega) \pm \sigma_{\nu\mu}(\omega)}{2}.
\end{align}
From now on, we will refer to the symmetric part $\sigmaS$ as longitudinal optical conductivity and the antisymmetric part $\sigmaH$ as optical Hall conductivity. The real part of the longitudinal optical conductivity $\Re \sigmaS$ 
determines the absorption of linearly polarized light, while %
the imaginary part of the optical Hall conductivity $\Im \sigmaH$  represents the differential absorption of left- and right-handed circularly polarized light, known as magnetic circular dichroism. %
These two components together constitute the absorptive (Hermitian) part of optical conductivity related to the energy dissipation: $\sigma^{\ab} \equiv \Re\sigmaS + i \Im\sigmaH= (\sigma+ \sigma^\dagger)/2$. $\sigma^{\ab}$ in insulators is given by: 
\begin{align}
    \sigma^{\ab}_{\mu\nu}(\omega) = \pi\omega e^2\int[\dd{\vec{k}}] \sum_{a, b} \delta(\ene_{ba}-\hbar\omega) A^{\mu}_{ab} A^{\nu}_{ba} f_{ab}. \label{eq:sigma_abs_T}
\end{align}

To establish a direct connection between optical absorption and quantum geometry, we introduce a generalized optical weight,  
\begin{eqnarray}
W^n_{\mu\nu}(\Omega) &\equiv& \int_0^{\Omega} \dd\omega \frac{ \sigma^{\ab}_{\mu\nu}(\omega)}{\omega^n},  \label{eq:generalizedSW}
\end{eqnarray}
where $n\geq 0$ is an integer. $W^{0}(\Omega)$ is the standard optical spectral weight below a cutoff frequency $\Omega$, with $\Omega=\infty$ corresponding to the full spectral weight. 
$W^{n\geq 1}$ can be regarded as the negative-$n$-th moment of $\sigma^{\ab}$ in frequency domain. For insulators, since optical absorption only occurs at frequencies above the gap, the integral in Eq.~\eqref{eq:generalizedSW} is convergent and the generalized spectral weight $W^n$ is finite for all $n$.  For any given $n$, $W^n(\Omega)$ as a function of the cutoff frequency $\Omega$ contains the full information of optical conductivity, as $\sigma^{\ab}(\omega) = \omega^n \dv*{W^n}{\omega}$ and the reactive (anti-Hermitian) part of the optical conductivity $\sigma^{\re}(\omega) = \Re\sigmaH + i\Im\sigmaS$ can be further obtained from $\sigma^{\ab}$ through the Kramers-Kronig relation.

In this work, we focus on the negative-first moment of the absorptive part of optical conductivity $W^1(\Omega)$. %
From Eq.(\ref{eq:sigma_abs_T}) we have     
\begin{eqnarray}
W^1_{\mu \nu}(\Omega) &\equiv& \int_0^{\Omega} \dd\omega \frac{ \sigma^{\ab}_{\mu\nu}(\omega)}{\omega}  \nonumber \\
&=& \frac{\pi e^2}{\hbar} \sum_{c, v}  \int_{\ene_{cv} \leq \Omega}[\dd{\vec{k}}] A^{\mu}_{vc} A^{\nu}_{cv}. \label{eq:loss}
\end{eqnarray}
where $c, v$ run over the conduction bands and the valence bands respectively, with $f_v=1$ and $f_c=0$. 
For a given pair of conduction and valence bands, the $\vec k$ integral extends over the region where the energy gap $\ene_{cv}$ is below $\hbar \Omega$.    
Since $\Re \sigma^{\ab}(\omega)/\omega = \epsilon''(\omega)$ is the imaginary part of the complex dielectric constant known as the loss factor, 
the real part of $W^1(\Omega)$ represents the integrated dielectric loss below frequency $\Omega$. 

In the limit $\Omega\rightarrow \infty$, $W^1(\infty)$ receives contributions from all interband transitions. Then,  the $\vec k$ integral in Eq.~\eqref{eq:loss} extends over the whole Brillouin zone, and $c, v$ run through all conduction and valence bands respectively. In this case, the right-hand side of Eq.~\eqref{eq:loss} only depends on the interband Berry connection, which 
suggests a quantum geometric origin of the generalized spectral weight integrated over all frequencies $W^1(\infty)$.  

More broadly, optical absorption below a given cutoff frequency $\Omega$ involves a finite number of bands around the Fermi level. For the moment, suppose that the energy gaps between  $m$ highest valence bands and $n$ lowest conduction bands, $\ene_{cv}(\vec k)$, is smaller than $\hbar \Omega$ at all $\vec{k}$, while the energy gap to any higher conduction bands exceeds $\hbar \Omega$. Then, optical absorption below frequency $\Omega$ comes entirely from interband transitions among these $m+n$ bands over the whole Brillouin zone, and  
we have 
\begin{eqnarray}
\int_0^{\Omega} \dd\omega \frac{ \sigma^{\ab}_{\mu\nu}(\omega)}{\omega} 
= \frac{\pi e^2}{\hbar} \sum_{c=1}^m \sum_{v=1}^n  \int[\dd{\vec{k}}] A^{\mu}_{vc} A^{\nu}_{cv}. \label{eq:loss2}
\end{eqnarray}
This expression only involves interband Berry connection as well, calling for a quantum geometric understanding.   

\section{Quantum Geometry and Generalized $f$-sum rule}  \label{sec:geometry_fsum_rule}

To develop a quantum geometric theory of interband optical conductivity, it is necessary to consider multi-band manifold over $\vec{k}$ space. A set of bands $\{ \ket{u_{\vec{k}1}}, ..., \ket{u_{\vec{k}r}} \}$ in $k$-space defines a family of $r$-dimensional Hilbert subspace parameterized by the wavevector $\vec{k}$, i.e., a vector bundle of rank $r$ over the Brillouin zone. 
The geometry of this multi-band manifold can be characterized by a quantum geometric tensor, a $r\times r$ matrix $\gqmet^{\mu\nu}$ with matrix elements defined by (see for example~\cite{ma_abelian_2010})
\begin{align}
    \gqmet^{\mu\nu}_{ij} &= \mel{\partial_\mu u_i}{(1-P)}{\partial_\nu u_j} \quad \text{with $i,j = 1, ..., r$}, \label{eq:gqmet_ij_def} 
\end{align}
where $\proj =\sum_{i=1}^r \ket{u_i} \bra{u_i} $ is the projection operator associated with the $r$-dimensional subspace spanned by the bands of interest.  

In the single-band case ($r=1$), $\gqmet^{\mu\nu}$ reduces to a scalar---the Abelian quantum geometric tensor, whose symmetric part and anti-symmetric part with respect to the spatial indices $\mu, \nu$ are known as the quantum metric %
and the Berry curvature %
respectively. 
In the case of a multi-band manifold ($r>1$), $\gqmet^{\mu\nu}$ is a $r\times r$ matrix---the non-Abelian quantum geometric tensor. We can readily show $(\gqmet^{\mu\nu})^\dagger = \gqmet^{\nu\mu}$. Its symmetric part and antisymmetric part define the non-Abelian quantum metric and the non-Abelian Berry curvature respectively: 
$\gqmet^{\mu\nu} = \manifsm^{\mu\nu} -\frac{i}{2}\manicurv^{\mu\nu}$,  %
both of which are $r\times r$ Hermitian matrices.

Armed with the notion of the quantum geometric tensor for multi-band manifolds,  we now establish a direct connection between optical absorption and quantum geometry. First, note that
\begin{eqnarray}
\sum_{c,v} A^{\mu}_{vc} A^{\nu}_{cv} &=&  \sum_{v} \mel{\partial_\mu u_v}{P_c}{\partial_\nu u_v}  \nonumber \\
&=&\sum_{v}   \mel{\partial_\mu u_v}{(1-P_v)-(1-P_v - P_c)}{\partial_\nu u_v}   \nonumber 
\end{eqnarray}
where $P_{v,c}$ is the projection operator onto the valence/conduction band manifold. Using this identity, 
we can rewrite Eq.~\eqref{eq:loss}: 
\begin{align}
 \int_0^{\Omega} \dd\omega \frac{ \sigma^{\ab}_{\mu\nu}(\omega)}{\omega}  = \frac{\pi e^2}{\hbar}  \int[\dd{\vec{k}}]  \left( \Tr\gqmet^{\mu\nu}_v-\Tr_v\gqmet^{\mu\nu}_{0} \right), \label{eq:sum_rule_abs}
\end{align}
where $\gqmet_v, \gqmet_{0}$ are non-Abelian quantum geometric tensors associated with the $m$-dimensional valence band manifold and the $(m+n)$-dimensional manifold of combined valence and conduction bands, respectively. $\Tr_v(\dots)$ is the partial trace over the valence band, namely, $\Tr_vO \equiv \sum_v O_{vv}$.

Eq.~\eqref{eq:sum_rule_abs} shows that the optical absorption corresponds to the change of the quantum geometry of the subspace spanned by $m$ valence bands when the $n$ conduction bands are added. If there is no optical transition allowed between the valence and conduction bands due to, e.g., a symmetry-based selection rule, the quantum geometry of valence bands is unchanged when the Hilbert subspace is enlarged to include conduction bands: $A^{\mu}_{vc}=0$ leads to $\Tr\gqmet^{\mu\nu}_{v} = \Tr_v\gqmet^{\mu\nu}_{0}$.

We now further show that the real part of Eq.~\eqref{eq:sum_rule_abs}---the integral of $\Re \sigmaS_{\mu\nu}(\omega)/\omega$---can be expressed in an elegant form using the quantum metric tensors alone:
\begin{align}
    & \int_{0}^{\Omega} \dd{\omega}  \frac{\Re\sigmaS_{\mu\nu}(\omega)}{\omega}  \nonumber \\
    &= \frac{\pi e^2}{2\hbar}\int[\dd{\vec{k}}]  \qty(\sum_{v=1}^m\mel{\partial_{\mu} u_v}{P_c}{\partial_{\nu} u_v} + \sum_{c=1}^n\mel{\partial_{\mu} u_c}{P_v}{\partial_{\nu} u_c}) \nonumber \\
    &= \frac{\pi e^2}{2\hbar}  \int[\dd{\vec{k}}]  \left( \Tr\manifsm^{\mu\nu}_v +\Tr\manifsm^{\mu\nu}_c 
    -\Tr\manifsm^{\mu\nu}_0 \right). \label{eq:absorption_partial}
\end{align}
Here $\manifsm_v, \manifsm_{c}, \manifsm_0$ are the quantum metric tensor for the manifold of valence bands, conduction bands, and these bands combined, respectively. 

Eq.~\eqref{eq:absorption_partial}  is the first main result of this work. It directly relates the optical absorption to the trace of the non-Abelian quantum metric tensors. For a multi-band manifold, $\Tr\manifsm^{\mu \nu}$ is a positive semi-definite $d\times d$ matrix that is independent of the choice of basis states. It measures how the multi-band subspace changes with $\vec{k}$, which can be seen from the squared norm of the change of the projection operator $P$ to second order in $\vec{k}$: 
\begin{eqnarray}
\Tr[(\delta P)^2] = \sum_{\mu,\nu} 2\Tr \manifsm^{\mu\nu} \delta k_\mu \delta k_\nu. \label{deltaP}
\end{eqnarray}
Note that the quantum metric of a manifold is {\it not} the sum of the ones for its subspaces. In fact, the equation~\eqref{eq:absorption_partial} shows that the generalized optical weight precisely measures the difference of the two.

As an example, let us consider a two-dimensional free electron gas under a magnetic field, which exhibits Landau levels equally spaced by cyclotron energy $\hbar \omega_c$ with $\omega_c = \abs{eB}/m$. 
An optical transition only occurs between two adjacent Landau levels $n-1 \leftrightarrow n$ due to the angular momentum selection rule. Based on Eq.~\eqref{eq:absorption_partial}, we can use the quantum metric tensors to calculate generalized optical weight $W^1$ associated with this inter-Landau-level transition. 
It is straightforward to show that the Abelian quantum metric tensor for $n$-th Landau level is constant in $\vec k$ space: $g^{\mu \nu}_n = (\hbar/\abs{eB})(n+1/2) \delta^{\mu \nu}$, while the trace of the non-Abelian quantum metric tensor for the two-Landau-level manifold is given by $\Tr \manifsm_0^{\mu \nu} = (\hbar /\abs{eB}) n \delta^{\mu \nu}$. Then, for the integer quantum Hall state at filling factor $\nu$,  we have 
\begin{eqnarray}
\int_{0}^{\Omega} \dd{\omega}  \frac{\Re\sigma_{ii}(\omega)}{\omega} =\frac{e^2}{\hbar} \frac{\nu}{4} \; \textrm{for } \Omega>\omega_c.  \label{LL}
\end{eqnarray}
This result agrees with a direct calculation of optical conductivity.

Returning to the discussion on general systems, in the limit of the cutoff frequency $\Omega \rightarrow \infty$, an interband optical transition is allowed between any pair of occupied and unoccupied bands. In this case, our general formula Eq.~\eqref{eq:sum_rule_abs} and Eq.~\eqref{eq:absorption_partial} can be further simplified. Since the complete set of bands spans the entire Hilbert space, the corresponding quantum geometric tensor $\gqmet_0^{\mu \nu}=0$ so that Eq.~\eqref{eq:sum_rule_abs} reduces to 
\begin{align}
    \int_0^{\infty} \dd{\omega} \frac{\sigma^{\ab}_{\mu\nu}(\omega)}{\omega} &= \frac{\pi e^2}{\hbar} \int[\dd{\vec{k}}] \Tr\gqmet^{\mu\nu}. \label{eq:optical_sum_QGT} 
\end{align}
This elegant formula directly relates the negative-first moment of optical conductivity $W^1(\infty)$ to the quantum geometry of the ground state wavefunction. We regard Eq.~\eqref{eq:optical_sum_QGT} as a generalized optical sum rule complementary to the standard $f$-sum rule which relates the optical spectral weight $W^0(\infty)$ to the charge stiffness (Drude weight).

We now discuss the real and imaginary parts of Eq.~\eqref{eq:optical_sum_QGT} separately. The imaginary part relates magnetic circular dichroism $\sigmaH$ to %
the Chern invariant of the ground state:
\begin{align}
    \int_0^{\infty} \dd{\omega} \frac{\Im\sigmaH_{\mu\nu}(\omega)}{\omega} &= -\frac{e^2}{4\hbar} C^{\mu\nu}. \label{eq:dichroism}
\end{align}
where the Chern invariant $C^{\mu \nu} \equiv 2\pi \int[\dd{\vec{k}}] \sum_v \Omega^{\mu\nu}_v$ is a topological invariant defined by the integral of the total Berry curvature of the occupied bands over the Brillouin zone. In two dimensions, $C^{\mu \nu} \equiv \epsilon^{\mu \nu} C$ where $C$ is an integer. In three dimensions, $C^{\mu\nu} \equiv \epsilon^{\mu \nu \lambda} C_\lambda$ where $C_\lambda$ is a reciprocal lattice vector. 

It is well known since the work of Thouless-Kohmoto-Nightingale-Nijs (TKNN)~\cite{Thouless1982} that the Chern invariant of an insulating state manifests in quantized dc Hall conductance: 
$\sigma^H = C  e^2/h$ in two dimensions and $\sigma^H_{\mu \nu} = \epsilon_{\mu \nu \lambda} C_\lambda e^2/h$ in three dimensions.   
The quantized Hall effect is a dissipationless transport phenomenon accompanied by zero longitudinal resistivity. Alternatively, Eq.~\eqref{eq:dichroism} shows that the Chern invariant can be measured directly by magnetic circular dichroism, the difference in optical absorption of left- and right-handed circularly polarized lights. 

It is remarkable that the Chern invariant can be measured through both dc Hall transport and optical magnetic circular dichroism. This is not a coincidence. Eq.~\eqref{eq:dichroism} and the TKNN formula are directly related through the Kramers-Kronig relation between the absoprtive and reactive parts of optical conductivity:  $
\frac{i}{\pi}P\int_{-\infty}^\infty \dd{\omega} \sigma^{\ab}_{\mu\nu}(\omega)/(\omega-\omega') = \sigma^{\re}_{\mu\nu}(\omega')
$, 
where $\sigma^{\re}(\omega) = \Re\sigmaH + i\Im\sigmaS$. 
By setting $\omega'=0$ and using the general property $\sigma^{\ab}(-\omega)=(\sigma^{\ab}(\omega))^{*}$, the Kramers-Kronig relation connects magnetic circular dichroism to dc Hall conductivity:   
   $\int_{0}^\infty \dd{\omega} \Im \sigma^{H}_{\mu\nu}(\omega)/\omega = -\frac{\pi}{2} \sigma^{H}_{\mu\nu}(0)$.  
For isotropic optical conductivity, this reduces to Eq.~\eqref{eq:KK} in the introduction.

From now on, we focus on the real part of Eq.~\eqref{eq:optical_sum_QGT}:     
\begin{align}
    \int_0^{\infty} \dd{\omega} \frac{\Re\sigmaS_{\mu\nu}(\omega)}{\omega} &= \frac{\pi e^2}{\hbar} \int[\dd{\vec{k}}] g^{\mu\nu}. 
 \label{eq:absorption}
\end{align}
where $g^{\mu \nu} \equiv \Tr\manifsm^{\mu \nu}$ is the trace of the non-Abelian quantum metric tensor of the occupied band manifold. Equivalently,  $g^{\mu\nu}$ is  Abelian quantum metric tensor of the Slater determinant state made of all the occupied bands: $|u_{\vec{k}1} u_{\vec{k}2} ... u_{\vec{k}r}|$.  

We note that Eq.~\eqref{eq:absorption} was first derived by a different method in the early seminal work of Souza, Wilkens and Martin on electronic polarization and localization in insulators \cite{souza_polarization_2000}. There, the quantum metric is defined for the many-body ground state with a twisted boundary condition. In the case of noninteracting band insulators, their result reduces to Eq.~\eqref{eq:absorption}. We also note that the integral of the trace of the quantum metric is directly related to the spread of the Wannier function in real space, as shown in the pioneering work by Marzari and Vanderbilt~\cite{marzari_maximally_1997}. 

While it has existed for a quarter of a century, the implication of Eq.~\eqref{eq:absorption} for optical conductivity and quantum geometry is not adequately explored. Some studies in this direction can be found in ~\cite{souza_dichroic_2008, Neupert2013, ahn_riemannian_2022} and references therein. We hold the view that  Eq.~\eqref{eq:absorption} is a fundamental relation between the optical absorption and the ground state property, which is universally applicable to {\it all} insulators. %

The right-hand side of Eq.~\eqref{eq:absorption}---the integral of the quantum metric of the occupied bands $g^{\mu \nu}$ over the Brillouin zone---is a quantum property of insulating ground states. 
Because of its relation to generalized optical weight, we call it ``quantum weight'':   
\begin{eqnarray}
 K^{\mu\nu} \equiv 2\pi \int[\dd{\vec{k}}] g^{\mu\nu}.
\end{eqnarray}
As we shall demonstrate below, the quantum weight is a central quantity that links together band topology, optical absorption and the insulating gap.

\section{Infrared absorption near topological band inversion} \label{sec:infrared_absorption}

The quantum weight provides a quantitative measure of the degree of ``quantumness'' in the insulating state. 
To illustrate this point, we consider two distinct types of insulators: atomic insulators and topological insulators, which have small and large quantum weight respectively.  

In an atomic insulator, electrons occupy highly-localized atomic orbitals $\phi_n(\vec{r}-\vec{R})$ located at lattice sites $\vec{R}$. The characteristic size of these orbitals $\xi$ is small compared to the lattice constant $a$, hence there is no hopping between sites. 
In this case, the Bloch wavefunction $\psi_{n\vec{k}}(\vec{r})$ is given by:   
$  
\psi_{n\vec{k}}(\vec{r}) = (1/\sqrt{N})\sum_{\vec{R}}e^{i\vec{k}\vdot\vec{R}}\phi_n(\vec{r}-\vec{R}).
$ 
Assuming that the spatial overlap between atomic orbitals on different sites is negligible, it is straightforward to show that the quantum geometric tensor of the occupied band manifold is related to the matrix elements of position operator between atomic orbitals on the same site:  
$   
\gqmet^{\mu\nu}_{ij} =\manifsm^{\mu\nu}_{ij} = \sum_{n} \mel{\phi_i}{r^{\mu}}{\phi_n} \mel{\phi_n}{r^{\nu}}{\phi_j} 
$
where $i,j$ belong to occupied orbitals, and $n$ run through unoccupied orbitals. Therefore the quantum weight $K \sim (\xi/a)^2 a^{2-d}$ ($d$ is spatial dimension) is also small, resulting in weak optical absorption.    

The opposite case of large quantum weight can be found in narrow gap insulators near topological band inversion. When inversion symmetry is present, the effective Hamiltonian  $H(\vec{k})$ for low-energy states  generally takes the forms of a massive Dirac fermion~\cite{fu_topological_2007}: %
\begin{align}
    H(\vec{k}) = %
     \Delta \Gamma_0 + v \sum_{\mu=1}^d k_\mu \Gamma_{\mu}  \label{eq:Dirac} 
\end{align}
where $\Gamma_0, ..., \Gamma_d$ are $4\times 4$ Dirac Gamma matrices satisfying $\{ \Gamma_i, \Gamma_j \} =2\delta_{i j}$. Tuning $\Delta$ from positive to negative induces a band inversion at $\vec{k}=0$ and results in a phase transition between topologically distinct insulators. At the critical point $\Delta=0$, the low-energy spectrum is described by massless Dirac fermions.    

We now calculate the quantum metric $g$ and quantum weight $K$ for this system, and evaluate the generalized optical weight $\Sigma$. When $H(\vec k)$ takes the form 
\begin{eqnarray}
H(\vec k) = E_{\vec k} \sum_{\lambda=0}^d n^\lambda_{\vec k } \Gamma_\lambda     \label{Hk}
\end{eqnarray} 
with $E_{\vec k}>0$ and $\vec n$ is a  unit vector, the projection operator for the occupied bands can be written as 
$P_{\vec k}=  \frac{1}{2}(1+ \sum_\lambda n^\lambda_{\vec k} \Gamma_\lambda)$. 
Using Eq.~\eqref{deltaP}, we find the quantum metric tensor is 
\begin{eqnarray}
g^{\mu \nu} = \frac{1}{2}(\partial_\mu \vec{n}) \cdot (\partial_\nu \vec{n}). 
\end{eqnarray}

Provided that the system has an insulating gap, $\vec{n}$ is well defined in $\vec k$ space, and 
the quantum metric $g^{\mu\nu}$ is finite.  
Near the band inversion transition, however, $\vec{n}$ changes rapidly around ${\vec k}=0$ where the gap is small, leading to a large $g^{\mu \nu}$ that may give dominant contribution to the quantum weight.  
Indeed, for the Dirac Hamiltonian Eq.~\eqref{eq:Dirac}, we find ${\vec n}_{\vec k} = (\Delta, v \vec{k})/\sqrt{\Delta^2 + v^2 k^2}$. The trace of quantum metric tensor is 
\begin{eqnarray}
g_{\vec k} \equiv \sum_\mu g_{\vec k}^{\mu \mu} =  \frac{1}{2} \frac{(d-1) v^2 k^2 + d \Delta^2}{(v^2 k^2 + \Delta^2)^2}. 
\end{eqnarray}
As $\Delta\to 0$, $g$ diverges as $1/k^2$ near $\vec{k}=0$. Therefore, near the topological phase transition, the quantum weight has non-analytic dependence on the insulating gap $\Delta$ due to the contribution from low-energy states: $K^{\mu \nu} = (K/d) \delta_{\mu\nu}$ where the asymptotic form of $K$ at small $|\Delta|$ is given by  
\begin{eqnarray}
K  \sim  & | \Delta |  &  (d=3)  \nonumber \\
&  \log(|\Delta|) & (d=2) \nonumber \\ 
&  1/|\Delta| & (d=1) \label{K-Delta}
\end{eqnarray} 
Importantly, the quantum weight exhibits a logarithmic divergence in two dimensions and a power-law divergence in one dimension.

By the general relation Eq.~\eqref{eq:absorption} between quantum weight and generalized optical weight, a divergent quantum weight necessarily implies strong optical absorption at low photon energy. 
Indeed, a well-known example is two-dimensional massless Dirac fermion systems such as graphene. Here, the real part of optical conductivity takes the universal value $\Re \sigma = \pi e^2/(2h)$ over a broad range of frequencies. Therefore, the negative-first moment of optical conductivity, $\int d\omega  \Re\sigma/\omega$,  has a logarithmic divergence at low energy, which leads to the $\log |\Delta|$ dependence in the presence of a Dirac mass gap, matching the quantum weight shown by Eq.~\eqref{K-Delta}.  

Unlike the standard spectral weight $\int d\omega \Re \sigma $ which gives equal weight to low and high frequency, the generalized optical weight $\int d\omega \Re \sigma/\omega$ gives large weight to optical conductivity at low frequency. By $f$-sum rule, the full spectral weight only depends on the electron density and therefore is insensitive to any details of the system. In contrast, 
the generalized optical weight is directly connected to the ground state wavefunction as shown by  Eq.~\eqref{eq:absorption}, and therefore 
provides a powerful tool for studying topological phase transitions involving a dramatic change in the ground state, such as the topological band inversion discussed above.

\section{Quantum weight, topological invariant, and 100\ \% magnetic circular dichroism. } \label{sec:weight_invariant_dichroism}

We have so far established a direct relation between optical absorption and quantum geometry. In particular, the generalized optical weight integrated over all frequencies is given by the quantum weight of the occupied bands. In this section, we consider quantum weight of topological bands. We mainly focus on Chern insulators in two dimensions, noting that the generalization to three dimensions and to quantum spin Hall insulators with conserved spin $U(1)$ symmetry is straightforward.

The quantum weight in two dimensional systems is a dimensionless quantity.  %
Interestingly, the quantum weight is lower bounded by the topological Chern number. This lower bound naturally follows from the relation between the quantum geometry and optical absorption as shown below. The crucial observation is that, the optical absorption, which is related to quantum geometry, is always non-negative. 

Let us consider the case of a circularly polarized light at frequency $\omega$: $E_x(t)= E \cos (\omega t)$ and $E_y = \pm E \sin (\omega t)$, or equivalently, ${\vec E}_{\omega} = E (\hat{x} \pm i \hat{y})$, with $\pm$ corresponding to left and right handedness. The induced current is given by ${\vec j}_\omega = (\sigma_{xx} \pm i \sigma_{xy},\sigma_{yx} \pm i \sigma_{yy}) E$.  The absorbed power is thus 
\begin{eqnarray}
\Re ( {\vec j}^*_\omega  \cdot {\vec E}_\omega) =  
\left(\Re(\sigma_{xx} + \sigma_{yy}) \pm  2 \Im \sigma^H_{xy}  \right) E^2.  \label{eq:absorbed_power}
\end{eqnarray} 
Since the absorbed power must be non-negative for every frequency, we must have
\begin{eqnarray}
\Re(\sigma_{xx} + \sigma_{yy}) \geq 2 \abs{\Im \sigma^H_{xy}}. \label{eq:sigma_bound}
\end{eqnarray} 
This is the relation~\eqref{eq:abs_inequality} we used in the introduction.
By dividing $\omega^n$ and integrating over frequency, we obtain a general inequality between the moments of $\sigma^{\ab}$ as
\begin{align}
    W^n_{xx}(\Omega) + W^n_{yy}(\Omega) \geq 2 \abs{W^n_{xy}(\Omega)}. 
\end{align}
In particular, for $n=1$ and $\Omega=\infty$, we can rewrite this with the quantum weight $K$ and the Chern number $C$ with Eq.~\eqref{eq:optical_sum_QGT} and ~\eqref{eq:dichroism}. Then we obtain the following inequality between the quantum weight and Chern number $C$:
\begin{align}
     K \equiv \sum_{i=x,y} K^{ii} \ge \abs{C}, 
    \label{eq:bound_on_quantum_weight}
\end{align}
where $C \equiv 2\pi  \int[\dd{\vec{k}}]  \Omega^{xy}$ is the Chern number of the ground state. It is clear from our derivation above that when the quantum weight of the occupied band manifold saturates the Chern number bound $K=\abs{C}$, the equality in Eq.~\eqref{eq:sigma_bound} must also be satisfied at all frequencies---that is to say, the system only absorbs circularly polarized light of one handedness, but not the other at all, namely, exhibits \SI{100}{\percent} magnetic circular dichroism.

The lower bound on the quantum weight~\eqref{eq:bound_on_quantum_weight} can also be derived from an inequality between the quantum metric and the Berry curvature: $\sqrt{\det g} = \sqrt{g_{xx}g_{yy}-g_{xy}g_{yx}}\ge \abs{\Omega^{xy}}/2$. This inequality was first derived for single band cases by Roy~\cite{roy_band_2014} and later generalized to multiple band cases~\cite{peotta_superfluidity_2015}. Noting that $\tr g=(g_{xx}+g_{yy}) \ge 2\sqrt{\det g}$ and integrating the inequality over the Brillouin zone, 
we recover the bound~\eqref{eq:bound_on_quantum_weight}. It is remarkable that the mathematical relation between the quantum metric and the Berry curvature is closely linked to the fact that the optical absorption of circularly polarized light must always be non-negative. 

Note that the quantum metric $g$ and the Berry curvature $\Omega$ used here are for the Slater determinant state of the entire occupied band manifold, $|u_{\vec{k}1} u_{\vec{k}2} ... u_{\vec{k}r}|$, which applies to an arbitrary number of occupied bands. Also note that while the Chern number is additive, i.e., the Chern number $C$ for the occupied band manifold is the sum of the Chern number for each band, the quantum weight $K$ is not.

As an example, consider a two-band Hamiltonian of the form shown in Eq.~\eqref{Hk}, with $d=2$ and Gamma matrices replaced by $2\times 2$ Pauli matrices $\sigma_x, \sigma_y, \sigma_z$. Then, the unit vector $\vec n_{\vec k}$ in $\vec k$ space defines a mapping from the two-dimensional Brillouin zone (which is a torus) to the Bloch sphere.  The Chern number $C$ and the quantum weight $K$ are given by:  
$
C=\frac{1}{2}\int[\dd{\vec{k}}] \; {\vec n} \cdot (\partial_{x} {\vec n} \times \partial_{y} {\vec n})
$, %
$
K= \frac{1}{4} \int [\dd{\vec{k}}] \; \sum_{\mu=x,y}(\partial_{\mu} {\vec n})^2,
$
respectively. Then, from the inequality $(1/2)\sum_{\mu=x,y}(\partial_{\mu} {\vec n})^2 \geq |{\vec n} \cdot (\partial_{1} {\vec n} \times \partial_{2} {\vec n}) |$, the bound $K\geq |C|$ follows immediately. 
This bound is saturated when $\vec{n}_{\vec{k}}$ takes special instanton configurations~\cite{jian_momentum-space_2013}. 

We may call Chern insulators having minimum quantum weight $K=\abs{C}$ ``minimal Chern insulator''. An example is the integer quantum Hall state in Landau level systems: for {\it any} integer filling $\nu\geq 1$, it can be shown that the quantum weight of occupied Landau level manifold is  $K=\nu=\abs{C}$. It should be noted that the concept of minimal Chern insulator applies to systems with an arbitrary number of occupied bands. In multiband cases, $C$ and $K$ are defined through the non-Abelian quantum geometric tensor. In the special case of a single occupied band where $C$ and $K$ are defined by the Abelian quantum geometric tensor, the condition for the minimal Chern insulator $K=\abs{C}$ implies two conditions: (1) the so-called trace condition for the Chern band, $\tr g=\abs{\Omega^{xy}}$, is satisfied at every $\vec k$ point, and (2) either $\Omega^{xy}$ or $-\Omega^{xy}$ is positive semi-definite over the entire Brillouin zone~\cite{claassen_position-momentum_2015, wang_exact_2021, ledwith_vortexability_2022}.

In time-reversal-invariant systems with spin-orbit coupling and spin $s_z$ conservation, the Chern number of all occupied bands must be zero, but occupied spin-$\uparrow$ bands and spin-$\downarrow$ bands can have equal and opposite Chern numbers: $C_\uparrow=-C_\downarrow\equiv C_s$. In this case, the quantum weight is bounded by twice the spin Chern number: $K=K_\uparrow + K_\downarrow \geq |C_\uparrow| + |C_\downarrow| = 2 |C_s|$.

\section{Quantum weight and topological gap} \label{sec:weight_gap}
Next, we establish a general upper bound on the quantum weight of real materials: 
\begin{align}
 K \le \frac{ 2\pi \hbar^2 n}{m E_g}, \label{eq:bound_on_quantum_weight2}
\end{align}
with $n$ the electron density, $m$ the electron mass, and $E_g$ the energy gap of the insulator (the precise definition of $E_g$ will be discussed below). This inequality applies very broadly to systems whose Hamiltonian takes the form 
\begin{eqnarray}
H = \frac{{\vec p}^2}{2m}  + V(\vec r) +  {\vec p} \cdot {\vec A}(\vec r) 
+ {\vec A}(\vec r) \cdot {\vec p},    \label{eq:Hp2}
\end{eqnarray}
where %
$V$ and $\vec A$ can be any function of particle coordinate. Moreover, our result still holds when $V(\vec r)$ and $\vec A(\vec r)$ are spin (or pseudospin) dependent matrices.  %

We now derive the inequality relating quantum weight and energy gap, Eq.~\eqref{eq:bound_on_quantum_weight2}, from the perspective of optical absorption. We first note that optical absorption in insulators only occurs at frequencies above a threshold $\omega\ge E_g/\hbar$, where $E_g$ is the minimum energy required to optically excite the system, called the optical gap. $E_g$ must be greater than or equal to the gap in the energy spectrum denoted as $\Delta$: $E_g \geq \Delta$. For clean, noninteracting insulators, $E_g$ is the minimum direct band gap at which the optical transition is allowed.    
Since the real part of the optical conductivity $\Re\sigma_{ii}(\omega)$ onsets above $E_g$ and is always non-negative, 
 the negative-first moment of optical conductivity has an upper bound 
\begin{eqnarray}
\int^{\infty}_{0} d\omega \frac{\Re \sigma_{ii}(\omega)}{\omega} \leq 
\frac{\int^{\infty}_{0} d\omega \Re \sigma_{ii}(\omega)}{E_g/\hbar}.  \label{eq:inequalitySW}
\end{eqnarray}
By the standard $f$-sum rule~\cite{Kubo1957a}, %
when the Hamiltonian takes the form of Eq.~\eqref{eq:Hp2},  
the full optical spectral weight  is given by the charge stiffness: 
\begin{eqnarray}
\int^{\infty}_{0} d\omega \Re \sigma_{ii}(\omega) = \frac{\pi}{2} \frac{n e^2}{m} \label{eq:sum_rule}
\end{eqnarray}
which is independent of any details of the system.  
Combining Eqs.~\eqref{eq:inequalitySW},  ~\eqref{eq:sum_rule} and Eq.~\eqref{eq:absorption} immediately yields the upper bound on the quantum weight Eq.~\eqref{eq:bound_on_quantum_weight2}.

We further offer a heuristic argument for the inequality between quantum weight and energy gap of an insulator, Eq.~\eqref{eq:bound_on_quantum_weight2}.  
As shown in ~\cite{marzari_maximally_1997, souza_polarization_2000, resta_insulating_2011} (and references therein), the quantum weight $K$ is directly related to the electronic localization length $\xi$ in the insulating ground state: $K \sim (\xi/a)^2$ where $a$ is the lattice constant. On the other hand, the Heisenberg uncertainly principle dictates that for a given energy of confinement $E_g$, the localization length $\xi$ cannot be smaller than $\hbar /\sqrt{m E_g}$, as in a harmonic oscillator. This leads to the inequality between quantum weight and energy gap of the lowest band, which corresponds to the filling factor $\nu=1$ or density $n\sim 1/a^2$.

Putting together the lower and upper bounds on the quantum weight, Eq.~\eqref{eq:bound_on_quantum_weight} and ~\eqref{eq:bound_on_quantum_weight2}, we arrive at a remarkable relation
\begin{eqnarray}
|C| \leq K \leq \frac{ 2\pi \hbar^2 n}{m E_g}.  \label{eq:bound}
\end{eqnarray}
Eq.~\eqref{eq:bound} is a key result of our work. In one stroke, it links together band topology, quantum geometry, and energy gap of insulating states.

Interestingly, both lower and upper bounds on quantum weight are saturated simultaneously in the integer quantum Hall state of Landau level systems. Here, the energy gap is $E_g = \hbar eB/m$ and the density is $n= \nu B/\Phi_0$ where $\nu$ is the filling factor and $\Phi_0 = h/e$ is the flux quantum. Then, the upper bound $2\pi \hbar^2 n/(m E_g)=\nu$ is equal to the lower bound---the Chern number $C=\nu$. Then, the quantum weight must be $K=\nu$, matching our earlier result, Eq.~\eqref{LL}. That both bounds on the quantum weight are saturated is a special feature of Landau level systems.

The Landau level example shows that the lower and upper bounds on the quantum weight of insulators,   Eq.~\eqref{eq:bound}, are both tight for the general case.

\section{Microscopic and effective theories}
As our upper bound for quantum weight Eq.~\eqref{eq:bound_on_quantum_weight2} assumes the Hamiltonian of the form Eq.~\eqref{eq:Hp2}, we now discuss its applicability to real materials. As a matter of fact, the {\it microscopic} Hamiltonian for {\it all} solids takes the form of Eq.~\eqref{eq:Hp2}, with $m$ the bare electron mass and $V$ the periodic potential of the ions. 
Moreover,  the external magnetic field  and the microscopic spin-orbit interaction 
$\hbar/(4m^2 c^2) {\vec s} \cdot \nabla V \times \vec p$ %
can be captured by Eq.~\eqref{eq:Hp2} with spin-independent and spin-dependent vector potentials respectively. Therefore, the inequality Eq.~\eqref{eq:bound_on_quantum_weight2} constitutes a fundamental relation applicable to all real materials. Applied in this way, the mass and the carrier density in Eq.~\eqref{eq:bound_on_quantum_weight2} should be the bare electron mass $m_0$ and the total density of electrons including core electrons, in the same spirit that the optical spectral weight integrated over all frequencies counts all electrons.  

For practical purposes, we often use an effective Hamiltonian $H_{\mathrm{eff}}$, in the form of continuum or tight-binding model, to describe the low-energy degrees of freedom that are well separated from high-energy ones. %
For example, the effective theory of doped semiconductors is based on the $k\cdot p$ continuum Hamiltonian of doped electrons or holes with an effective mass. As another example, for tight-binding models, the effective Hamiltonian is described by a $k$-dependent matrix.
In these cases, the $f$-sum rule \eqref{eq:sum_rule} is modified as~\cite{Kubo1957a, hazra_bounds_2019}:  
    $\int  \dd{\omega}\sigmaS_{\mu\nu}(\omega) = \frac{\pi}{2} ne^2 (m_*^{-1})_{\mu\nu}, $
where the effective mass is given by 
\begin{align}
    (m^{-1}_*)_{\mu\nu} = \frac{1}{n}\int[\dd{\vec{k}}] \sum_{v} \mel{u_v}{(\partial_\mu\partial_\nu H_{\mathrm{eff},\vec{k}})}{u_v},
\end{align}
and $n$ is the carrier density in the effective theory. %

Since any effective theory and effective model parameters for solids are ultimately derived from the ``universal'' Hamiltonian Eq.~\eqref{eq:Hp2}, the fundamental bound with $n$ the total charge density and $m$ the bare mass always holds. Furthermore, a tighter bound can often be found within effective models.    
For example, consider a two-dimensional electron system with a weak periodic potential $V(\vec{r})$. The periodic potential induces the Bragg scattering of the electron wavefunctions and results in the formation of Bloch bands. In the low density limit $n a^2 \ll 1$ (where $a$ is the lattice constant and $n$ is total charge density), this periodic system is effectively described by a two-dimensional electron gas with an effective mass $m_*$. 
By treating the periodic potential as a perturbation, one can show that $m_*$ is always larger than the bare mass $m$. For example, for a square lattice potential $V(\vec{r})=2V(\cos(Gx) + \cos(Gy))$, the effective mass is given by 
$m_* = m\qty(1+\frac{8V^2}{\ene_G^2})$
with $\ene_G=\hbar^2 G^2/(2m)$. Now, consider applying an external magnetic field to create Landau levels. When the magnetic field is sufficiently weak, the cyclotron gap of the integer quantum Hall states is $\hbar eB/m_*$, which saturates the tighter bound given by the effective theory. In the opposite limit of a sufficiently strong field (beyond the applicable regime of the effective theory), the energy spectrum consists of free-electron Landau levels weakly perturbed by the periodic potential. In this case, the cyclotron gap is approximately $\hbar eB/m$, which saturates the fundamental bound.

\section{Topological gap bound} \label{sec:topological_gap_bound}

\begin{figure*}[htbp]
    \centering
    \includegraphics[width=1.9\columnwidth]{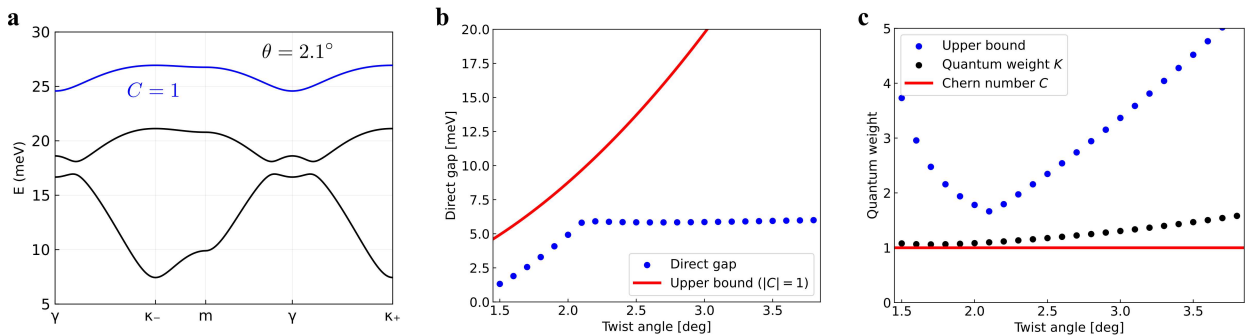}
    \caption{Topological gap bound in twisted homobilayer \MoTe. (a) The band structure of twisted \MoTe\ at twist angle $\theta=\ang{2.1}$. The blue line is the highest valence band that has the Chern number $C=1$. (b) The minimum direct gap between the highest and second-highest valence bands and the upper bound of the gap \eqref{eq:gap_bound} as a function of twist angle. (c) Quantum weight $K=K_{xx}+K_{yy}$, Chern number $C$, and the upper bound by the gap \eqref{eq:gap_bound} as a function of twist angle. As the gap, the minimum direct gap between the highest and second-highest valence bands is used.}
    \label{fig:MoTe2_bound}
\end{figure*}

Eq.~\eqref{eq:bound} also implies a remarkable relation of the topological gap and the Chern number:
\begin{align}
    E_g \le \frac{ 2\pi \hbar^2 n}{m \abs{C}}, \label{eq:gap_bound}
\end{align}
which provides an upper bound to the energy gap in Chern insulators. 
Note that the right-hand side depends only on the carrier density, the electron mass, and the Chern number of the ground state, and is independent of any details of the system. While the band topology describes the properties of the wavefunction rather than the energy dispersion, the relation \eqref{eq:gap_bound} shows that in real materials described by the Hamiltonian Eq.~\eqref{eq:Hp2}, the energy dispersion and the band topology cannot be completely independent. Our bound on the topological gap is saturated in Landau level systems as shown above, and therefore is tight for the general case. 

As seen from the above discussion,  Eq.~\eqref{eq:gap_bound} provides the upper bound on the optical gap $E_g$, i.e., the minimum energy required for optical absorption. Since the spectral gap $\Delta$ cannot exceed $E_g$, the bound~\eqref{eq:gap_bound} also applies to $\Delta$.

It is remarkable that our study of optical absorption from the perspective of quantum geometry has led to the discovery of the topological gap bound, a relation between band topology and energy spectrum which makes no reference to optical properties. Nonetheless, its connection to optical absorption is clear and direct. The upper bound on the topological gap can be reached if and only if optical absorption occurs at a single frequency $\omega=E_g/\hbar$, as shown by Eq.\eqref{eq:inequalitySW}.

Our bound \eqref{eq:gap_bound} is especially useful when applied to low carrier density systems with nontrivial band topology. As an example, we consider moir\'e superlattices formed from two-dimensional semiconductors. Recent theoretical calculations~\cite{wu_topological_2019, devakul_magic_2021} have shown that twisted homobilayer transition metal dichalcogenide (TMD) $t$MoTe$_2$ and $t$WSe$_2$ host topological bands over a broad range of twist angles. In particular, the topmost moir\'e valence bands in valley $K$ and $K'$, which carry opposite spins, have equal and opposite Chern numbers $C_\uparrow=-C_\downarrow=1$, resulting in a quantum spin Hall insulator at the filling of two holes per unit cell. The band topology is also manifested at the filling of one hole per unit cell, where the Coulomb interaction can induce spontaneous full spin/valley polarization and drive the system into a quantum anomalous Hall insulator \cite{devakul_magic_2021}. This state has been experimentally observed \cite{cai_signatures_2023,foutty_mapping_2023, zeng_integer_2023}.  

The moir\'e bands of $t$MoTe$_2$ or $t$WSe$_2$ are formed from the parabolic band of the monolayer by the presence of interlayer tunneling and layer-dependent potential that are spatially modulated by the moir\'e superlattice. Thus, the continuum model of moir\'e bands fits into the general form of Eq.~\eqref{eq:Hp2}. This allows us to apply Eq.~\eqref{eq:gap_bound}  to obtain an upper bound on the gap between the topmost and second moir\'e valence bands: 
\begin{eqnarray}
E^{\rm max}_g= \frac{2\pi \nu \hbar^2}{m_* A_\theta},    \label{Emax}
\end{eqnarray} 
with $\nu=1$ the filling factor for this case, $A_\theta = \sqrt{3} a_0^2/(2\theta^2)$ the area of the moir\'e unit cell, $a_0$ the monolayer lattice constant and $m_*$ the effective mass of holes in MoTe$_2$ or WSe$_2$.  We emphasize that this bound is completely independent of the form or strength of interlayer tunneling or superlattice potential. 

We compare the topological gap bound $E^{\rm max}_g$ with the minimum direct gap calculated from the continuum model of Ref.~\cite{wu_topological_2019}, using model parameters fitted to first-principles band structures for $t$MoTe$_2$~\cite{reddy_fractional_2023}.  Remarkably,  our bound is fairly tight at small twist angles as shown in Fig.~\ref{fig:MoTe2_bound}. At $\theta=2.1^\circ$,  the upper bound $E^{\rm max}_g=\SI{9.7}{\milli\electronvolt}$ is only $1.67$ times the calculated gap of \SI{5.8}{\milli\electronvolt}.

We further calculate the quantum weight $K$ using the Bloch wavefunction of the continuum model. Indeed, as shown in Fig.~\ref{fig:MoTe2_bound}, $K$ is found to be lower bounded by the Chern number $\abs{C}=1$ and upper bounded by  $E^{\rm max}_g /E_g$ where $E_g$ is the minimum direct gap in the continuum model.    
Remarkably, the quantum weight is very close to the Chern number throughout the twist angles shown here. On the other hand, its upper bound  $E^{\rm max}_g /E_g$ shows a deep minimum at $\theta= 2.1^\circ$. 

Our topological gap bound, Eq.~\eqref{Emax}, provides a key guiding principle for searching large gap topological insulators in two-dimensional electron systems. It is particularly useful for moir\'e materials, where first principles band structure calculation is challenging due to the large unit cell and strong lattice relaxation involved. Without relying on any microscopic details, we have shown that the topological gap is fundamentally limited by the average kinetic energy of charge carriers $2\pi \hbar^2 n /m^*$. This upper bound can only be increased by increasing the filling factor, reducing the moir\'e period,  or choosing materials with smaller effective mass. For $t$MoTe$_2$ with $m^*=0.62 m_0$, the topological gap between the first and second moir\'e bands cannot exceed \SI{35}{\milli\electronvolt} at $\theta=4^\circ$.   

Before concluding this section, we emphasize that Eq.~\eqref{eq:gap_bound} is quite general and holds even for interacting systems, as we showed in the introduction. In the next section, we will present a complementary understanding by formulating the quantum geometry for interacting systems.

\section{Interacting systems} \label{sec:generalization_interacting}

As shown in the introduction, the bound on the topological gap holds quite generally even in interacting systems. In this section, we show that all the main results of this work apply to interacting systems by constructing a many-body quantum geometric description of the optical conductivity. The key point is that both the standard $f$-sum rule Eq.~\eqref{eq:sum_rule} and our generalized sum rule Eq.~\eqref{eq:optical_sum_QGT} can be formulated for an interacting system in terms of the dependence of its ground state energy and wavefunction on the twisted boundary condition.

In order to describe the optical conductivity in interacting systems in terms of quantum geometry, we need to introduce the twisted boundary condition. 
We consider ground states with degeneracy $r$ in the $N$-electron system. Under twisted boundary condition, the ground states satisfies 
\begin{align}
    &\Psi_{a\vec{\kappa}}(\vec{r}_1, \dots, \vec{r}_i + \vec{L}_{\mu}, \dots, \vec{r}_N) \nonumber \\
    &=  e^{i\vec{\kappa}\vdot\vec{L}_{\mu}}\Psi_{a\vec{\kappa}}(\vec{r}_1, \dots, \vec{r}_i, \dots, \vec{r}_N), \label{eq:TBC}
\end{align}
where $\vec{L}_{\mu}=(0,\dots, L_{\mu}, \dots,0)$ specifies the system size in $\mu$-direction $L_{\mu}$, and the vector $\vec{\kappa}$ specifies the twisted boundary condition. 

The optical conductivity $\sigma(\omega;\vec{\kappa})$ for many-body systems can be calculated with the Kubo formula as a function of $\vec{\kappa}$. When the system is gapped, the optical conductivity does not depend on the boundary condition $\vec{\kappa}$ in the thermodynamic limit. Therefore, we can identify the optical conductivity with the one averaged over $\vec{\kappa}$, denoted by $\bar{\sigma}(\omega)$, as done in Ref.~\cite{niu_quantized_1985} for the dc Hall conductivity.  
Noting that the twisted boundary condition $\vec{\kappa}$ acts as the vector potential, $\bar{\sigma}(\omega)$ can be written as
\begin{align}
    &\bar{\sigma}_{\mu\nu}(\omega) = \frac{e^2}{\hbar} \int[\dd{\vec{\kappa}}] \sum_{n, m} \frac{-iE_{nm} A^{\mu}_{nm}A^{\nu}_{mn} f_{nm}}{\hbar\omega + E_{nm} + i\delta}, \label{eq:interacting_sigma}
\end{align}
where $E_{nm}=E_n-E_m$ is the energy difference between $n$-th and $m$-th many body eigenstate, and $\partial_{\mu}$ is the derivative with respect to $\kappa_{\mu}$. $f_{nm}=f_n-f_m$ with the probability $f_n$ that $n$-th eigenstate is realized. At zero temperature, the canonical distribution gives $f_n = 1/r$ when the state $n$ is one of the $r$-fold degenerated ground states and otherwise $f_n=0$.
$A^{\mu}_{nm}=\mel{n,\vec{\kappa}}{i\partial_{\mu}}{m,\vec{\kappa}}$ is the interband Berry connection for the $n$-th and $m$-th eigenstates the interacting system under the boundary condition $\vec{\kappa}$. As one can see from Eq.~\eqref{eq:interacting_sigma}, the expression is the same as the noninteracting cases~\eqref{eq:optical_conductivity} except that the quantum geometric quantities are defined with the boundary condition and that the band dispersion is replaced with the eigenenergy $E_n$. Therefore, we can relate them to the quantum geometry in the same way as the noninteracting cases.

The quantum geometric tensor $Q^{\mu\nu}$ for the many-body ground states is defined as 
\begin{align}
    \gqmet^{\mu\nu}_{ab} &= \mel{\partial_{\mu}\Psi_{a\vec{\kappa}}}{(1-P_{\vec{\kappa}})}{\partial_{\nu}\Psi_{b\vec{\kappa}}}.
\end{align}
where $\partial_{\mu}$ refers to the derivative with respect to $\kappa_{\mu}$, and $P_{\vec{\kappa}}$ is the projection operator onto the ground state subspace for the boundary condition $\vec{\kappa}$.
$\gqmet^{\mu\nu}$ is in general an $r\times r$ matrix and non-Abelian, satisfying $(Q^{\mu\nu})^{\dagger}=Q^{\nu\mu}$. The symmetric and antisymmetric components of $\gqmet^{\mu\nu}$ define the quantum metric and Berry curvature for the many-body states $\ket{\Psi_{\vec{\kappa}}}$ respectively: $\gqmet^{\mu\nu} = \manifsm^{\mu\nu} - i\manicurv^{\mu\nu}/2$. The trace of them gives the Abelian quantum geometry for the entire ground state subspace: $g^{\mu\nu}=\Tr\manifsm^{\mu\nu}, \Omega^{\mu\nu}=\Tr\manicurv^{\mu\nu}$. 
If the ground state is unique, the quantum geometric tensor reduces to the Abelian quantum geometric tensor.

Starting from the expression~\eqref{eq:interacting_sigma} and repeating the same procedure as the noninteracting cases, we can derive the bounds for many-body systems. Then we obtain relations for the interacting systems as: 
\begin{align}
    \abs{C} \le K \le \frac{2\pi\hbar^2 n}{mE_g},  \quad 
    E_g \le \frac{2\pi\hbar^2 n}{m\abs{C}}. \label{eq:manybody_gap_bound}
\end{align}
Here, $C$, $K$ are the Chern number and the quantum weight for one of the ground states defined as 
\begin{align}
    &K=2\pi\int[\dd{\vec{\kappa}}]\sum_{\mu, a} G^{\mu\mu}_{aa}f_a = \frac{K_{\mathrm{tot}}}{r}, \\
    &C=2\pi\int[\dd{\vec{\kappa}}]\sum_{a} F^{\mu\nu}_{aa}f_a = \frac{C_{\mathrm{tot}}}{r},
\end{align}
with the Chern number and the quantum weight for the entire ground state subspace $K_{\mathrm{tot}}=2\pi\int[\dd{\vec{\kappa}}]\sum_{\mu} g^{\mu\mu}, C_{\mathrm{tot}}=2\pi\int[\dd{\vec{\kappa}}]\sum_{a} \Omega^{\mu\nu}$.

The bound for many-body systems reduces to the inequalities~\eqref{eq:bound}, \eqref{eq:gap_bound} in noninteracting systems. Since the bounds for noninteracting systems are tight, the bounds for many-body systems are also tight in general. As in the noninteracting case, $E_g$ appearing in the bound is the optical gap. Since the spectral gap $\Delta$ is always larger or equal to $E_g$, the bound~\eqref{eq:gap_bound} holds for $\Delta$ as well, leading to Eq.~\eqref{eq:gapbound_intro} in the introduction.

We again emphasize that our results~\eqref{eq:manybody_gap_bound} hold quite generally. Our bound applies even to disordered or non-periodic systems. Remarkably the bound only depends on the charge density and the mass, and is completely independent of potential and interaction terms.    

As an example, consider band insulators in the presence of Coulomb interaction. In general, interaction-induced exciton effects reduce the optical gap, so that the upper bound of the energy gap~\eqref{eq:manybody_gap_bound} still holds.

Another important application of our topological gap bound is quantum Hall states in Landau level systems. It follows from Galilean invariance that the cyclotron resonance frequency is $\omega_c$  even in the presence of interaction~\cite{kohn_cyclotron_1961}; therefore, the optical gap $E_g=\hbar\omega_c$ always holds for Landau level systems at all filling factors. On the other hand, for the quantum Hall states with $\sigma_{xy}=Ce^2/h$, the filling factor $\nu$ is equal to $C$ for both integer and fractional quantum Hall states.  With $n=\nu eB/h$ and $\nu=C$, it follows that the optical gap bound~\eqref{eq:manybody_gap_bound} reduces to $E_g\le 2\pi \hbar^2 n/(m \abs{C})=\hbar\omega_c$. Therefore, the optical gap bound~\eqref{eq:manybody_gap_bound} is saturated for both the integer and fractional quantum Hall states. In this sense, Landau level systems are optimal as both integer and fractional Chern insulators. Our work further implies that even when a superlattice potential or any other perturbation violating Galilean invariance is added to Landau level systems, the optical gap still cannot exceed $\hbar \omega_c$. Finally, we note that the spectral gap $\Delta$ is considerably smaller than the optical gap $E_g$ in realistic fractional quantum Hall states. Indeed, the magneto-roton excitation significantly lowers $\Delta$ compared to $E_g = \hbar \omega_c$ in Landau level systems~\cite{girvin_magneto-roton_1986}.

In conclusion, we establish direct relations between three fundamental properties of insulators---the energy gap, quantum geometry, and topology---through the consideration of optical conductivity of solids. Our work %
opens new directions of research.

\begin{acknowledgements}
This work was supported by the U.S. Army Research Laboratory and the U.S. Army Research Office through the Institute for Soldier Nanotechnologies, under Collaborative Agreement Number W911NF-18-2-0048. YO is grateful for the support provided by the Funai Overseas Scholarship. 
LF was partly supported by the David and Lucile Packard Foundation. 
\end{acknowledgements}

\bibliography{references}

\end{document}